\documentclass[draftcls,onecolumn,12pt]{IEEEtran}
\usepackage{amssymb}
\usepackage{CJK}
\usepackage{amsmath}
\usepackage{amssymb,amsfonts}
\usepackage{indentfirst}
\usepackage[dvips]{graphicx}
\usepackage{float}
\usepackage{subfigure,color}
\usepackage{booktabs}
\usepackage{subfigure}
\begin{document}
\title{Multiple-Level Power Allocation Strategy for Secondary Users in Cognitive Radio Networks}
\author{\IEEEauthorblockN{Zhong Chen,  Feifei Gao,  Zhenwei Zhang,  James C. F. Li, and Ming Lei }
\thanks{Z. Chen and Z. W. Zhang are with  the 28th Research Institute of China Electronics Technology Group Corporation, Nanjing 210007, China.}
\thanks{
 F. Gao is with the
Tsinghua National Laboratory for Information Science and Technology,
Beijing, China, (e-mail: {zhong-chen08@mails.tsinghua.edu.cn,  feifeigao,
zxd-dau}@mail.tsinghua.edu.cn).}
\thanks{J. C. F. Li and M. Lei are
with NEC Laboratories China, Beijing, 100084, China. } }

\maketitle

\vspace{-10mm}
\begin{abstract}
In this paper, we propose a multiple-level power allocation
strategy for the secondary user (SU) in cognitive radio (CR)
networks. Different from the conventional strategies, where SU either
stays silent or transmit with a constant/binary power depending on
the busy/idle status of the primary user (PU), the proposed strategy
allows SU to choose different power levels according to a carefully
designed function of the receiving energy. The way of the power
level selection is optimized to maximize the achievable rate of
SU under the constraints of average transmit power at SU and average interference power at PU. Simulation results demonstrate that the proposed strategy can
significantly improve the capacity of  SU compared to the
conventional strategies.
\end{abstract}

\section{Introduction}
Cognitive radio (CR) has recently emerged as a promising technology
to improve spectrum utilization and to solve the spectrum scarcity
problem \cite{Godng2010}.  Consequently, spectrum sensing and power
allocation play as two key functionalities of a CR
system, which involves monitoring the spectrum usage and accessing the
primary band under given interference constraints.

The earliest spectrum access approach is the  \textit{opportunistic
spectrum access} where secondary user (SU) can only access the primary band when it
is detected to be idle \cite{Stotas2012}; The second approach is the
\textit{underlay} where SU is allowed to transmit beneath the primary user (PU) signal, and
sensing is not needed as long as the quality of service (QoS) of
PU is protected \cite{Gong2010}; The recent approach,
\textit{sensing-based spectrum sharing}, performs spectrum sensing
to determine the status of  PU and then accesses the primary band
with a high transmit power if PU is claimed to be absent, or with a
low power otherwise \cite{Fan2011,Kang20092}. These three approaches
adopt either constant or binary power allocation at SU which
is too ``hard'' and limits the performance of SU.

In this paper, we propose a multiple-level power allocation
strategy for SU, where the power level used at SU varies based on
its receiving energy during the sensing period. It can be easily
known  that the conventional constant or binary power allocations
are special cases of the proposed strategy. The whole strategy is
composed of: (i)~sensing slot, where the receiving energy is
accumulated  and the transmit power of SU is decided; (ii)
transmission slot, where SU sends its own data with the
corresponding power level. Different from the previous work \cite{Chen2013} where
 sensing and power allocation were studied for the scenario when PU transmits with
multiple power-level, in this paper, we consider PU transmits with constant power but
SU adopts multiple-level power.
 Under the constraints of the average
transmit power at SU and the average interference power at
PU, the sensing duration, energy threshold and power levels are
optimized  to maximize the average achievable rate at SU.

\section{System Model} \label{sectionreview}
\begin{figure}[h]
\centering
\includegraphics[width=90mm]{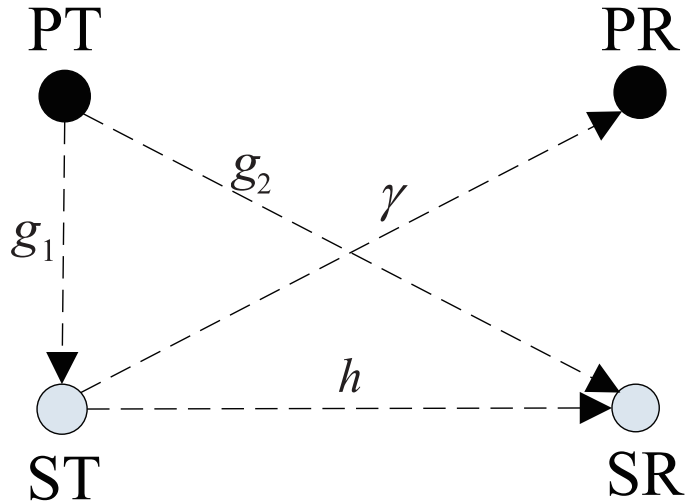}
\caption{System model of the cognitive radio network.}
\label{Systemmodel}
\end{figure}

Consider a CR network with a pair of primary and secondary
transceivers as depicted in Fig. \ref{Systemmodel}.
Let $g_{1}$, $g_{2}$, $\gamma$ and $h$ denote the instantaneous channel
power gains from the primary transmitter (PT) to the secondary
transmitter (ST),  from PT to the secondary receiver (SR), from ST
to the primary receiver (PR) and from  ST to SR, respectively.
We consider the simplest case that the channel gains are assumed to be constant and known at the secondary systems, since we focus on the proposed multiple-level power allocation strategy but not on the computing.
However, the idea and the results of the correspondence can be extended to other cases of full/ statistic/partial channel information.

One data frame of CR  is  divided into the sensing slot with
duration $\tau$ and the transmission slot with duration $T-\tau$.
During the sensing slot, ST listens to the primary channel and
obtains its accumulated energy. In the conventional schemes,
spectrum sensing is performed in this slot and the decisions on the
status (active/idle) of the channels are made. When transmitting, ST
accesses the primary band with the optimal power in order to
maximize the throughput while at the same time keeping the interference
 to PR.

During the sensing slot, the $j$th received sample symbol at ST is
\begin{equation}\label{TraP1}
    r_{j}=\left\{ \begin{aligned}
             &n_{j},& H_{0}, \\
             &\sqrt{g_{1}}s_{j}+n_{j},& H_{1},
                  \end{aligned} \right.
\end{equation}
where $H_{0}$ and $H_{1}$ denote the hypothesis that PT is absent
and present, respectively; $n_{j}$ is the additive noise
which is assumed to follow a circularly symmetric complex
  Gaussian distribution with zero mean and
variance $N_{0}$, i.e., $n_{j}\backsim{\cal N}_c(0, N_{0})$;
$s_{j}$ is the $j$th symbol transmitted
from PT. For the purpose of computing the achievable channel rate,
the transmitted symbols $s_{j}$ from the Gaussian constellation are typically assumed \cite{Fan2011, Kang20092},
i.e., $s_{j}\backsim {\cal N}_c(0, P_{p})$, where $P_{p}$ is the symbol power.  Without
loss of generality, we assume that $s_{j}$ and $n_{j}$ are
independent of each other.

The detection statistic $x$ using the accumulated received sample
energy  can be  written as
\begin{equation}\label{Xkafang}
   x=\sum_{j=1}^{\tau f_{s}}|r_{j}|^{2},
\end{equation}
where $f_{s}$ is the sampling frequency at ST. Then the probability
density functions (pdf), conditioned on $H_{0}$ and $H_{1}$, are
given by
\begin{equation}\label{eqjoieqwnt8}
\begin{aligned}
&f(x|H_{0})=\frac{x^{\tau f_{s}-1}e^{-\frac{x}{N_{0}}}}{\Gamma(\tau f_{s})N_{0}^{\tau f_{s}}},\\
&f(x|H_{1})=\frac{x^{\tau f_{s}-1}e^{-\frac{x}{N_{0}+g_{1} P_{p}}}}{\Gamma(\tau f_{s})(N_{0}+g_{1} P_{p})^{\tau f_{s}}},
\end{aligned}
\end{equation}
where $\Gamma(.)$ is the gamma function defined as $\Gamma(x)=\int_{0}^{+\infty}t^{x-1}e^{-t}dt$.

In the conventional CR, ST compares $x$ with a threshold
$\rho$, and makes  decision according to  $x \mathop
\gtrless\limits_{H_{0}}^{H_{1}}\rho$. Specifically:
\begin{itemize}
\item In opportunistic spectrum access approach,  ST can only
access the primary band when $x<\rho$ (it means $H_{0}$);

\item In sensing-based spectrum sharing, if $x<\rho$,  ST
transmits with one higher power and otherwise with a lower  power (binary power);

\item In underly approach,  ST transmits  with a constant
power for all $x$ according to the
interference constraint at PU (constant power). No sensing time slot is needed.
\end{itemize}

\section{Proposed Multiple-Level Power Allocation Strategy}
It can be easily realized that the conventional  constant or
binary power of SU does not fully exploit the capability of the
co-existing transmission. Motivated by this,  we propose a
multiple-level power allocation strategy for SU to improve the
average achievable rate.

\subsection{Strategy of Multiple-Level Power Allocation}\label{sectionproposedpower}
Define $\{\Re_{1},..., \Re_{M}\}$ as $M$ disjoint spaces of the
receiving energy $x$, and $\{P_{1},..., P_{M}\}$ as the
corresponding allocated power of SU. Then the proposed power
allocation strategy can be written as
\begin{align}\label{eqavethrou1}
      P(x)=\sum_{i=1}^{M}P_{i}I_{x\in \Re_{i}},
\end{align}
where $I_{A}$ is the indicating function that $I_{A}=1$ if $A$ is
true and $I_{A}=0$ otherwise. Note that, the conventional power
allocation rules are special cases when
 $M=1$ or $2$.

Using (\ref{eqavethrou1}), the instantaneous rates of  SU  with
receiving $x$, at the absence and the presence of PU, are given by
\begin{align}\label{eqaveth231rou}
     &R(x)|_{H_{0}}=\sum_{i=1}^{M}\textup{log}_{2}\left(1+\frac{P_{i}h}{N_{0}}\right)I_{x\in \Re_{i}},\\
     &R(x)|_{H_{1}}=\sum_{i=1}^{M}\textup{log}_{2}\left(1+\frac{P_{i}h}{N_{0}+g_{2}P_{p}}\right)I_{x\in \Re_{i}},
\end{align}
respectively. Then the average throughput of SU for the proposed
multiple-level power allocation strategy using the total probability formula can be formulated as
\begin{align}\label{eqavethrou}
R=&\frac{T-\tau}{T}\sum_{i=1}^{M}\left[q_{0}\textup{log}_{2}\left(1+\frac{P_{i}h}{N_{0}}\right)p_{i, 0}+q_{1}\textup{log}_{2}\left(1+\frac{P_{i}h}{N_{0}+g_{2}P_{p}}\right)p_{i, 1} \right],
\end{align}
where $q_{0}$ and $q_{1}=1-q_{0}$ are the idle and busy
probabilities of the PU respectively; $p_{i, 0} =\textup{Pr}(x\in
\Re_{i}|H_{0})$ and $p_{i, 1} =\textup{Pr}(x\in \Re_{i}|H_{1})$,
which  can be directly
computed from \eqref{eqjoieqwnt8} and are functions of $\tau$.

In order to keep the long-term power budget of SU, the average
transmit power, denoted by $\bar{P}$, is constrained as
\begin{eqnarray}\label{eq1c32}
\frac{T-\tau}{T}\sum_{i=1}^{M}P_{i}\left[q_{0}p_{i, 0} +q_{1}p_{i, 1} \right]\leq \bar{P}.
\end{eqnarray}

Moreover, to protect the QoS of PU, an interference temperature
constraint should be applied too.  Under (\ref{eqavethrou1}), the
interference is caused only when the PU is present. Denoting
$\bar{I}$ as the maximum average allowable  interference at PU,
 the average interference power constraint can be formulated as
\begin{eqnarray}\label{eq1c34}
    \frac{T-\tau}{T}\sum_{i=1}^{M}\gamma q_{1}P_{i}p_{i, 1}   \leq  \bar{I}.
\end{eqnarray}

Our target is to find the optimal space division
$\{\Re_{i}\}$,\footnote{Namely, we have multiple thresholds to
categorize $x$ rather than only using $\rho$ in convention.} the
power allocation $\{P_{i}\}$, as well as the sensing time $\tau$ in
order to maximize the average achievable rate of SU under the power
constraints. The optimization is then formulated  as
\begin{align}\label{OptP5789}
&\mathop\textup{max}\limits_{\tau, P_{i}, \Re_{i}}~R~\nonumber\\
&~~\textup{s.t}.~(\ref{eq1c32}),~(\ref{eq1c34}),~0\leq \tau \leq T,~P_{i}\geq 0,~\forall i.
\end{align}

The term $\frac{T-\tau}{T}$ means that the power constraints occur in the transmission slot.
Note that (\ref{OptP5789}) is nonlinear and non-convex over $\tau$. Hence, following \cite{Fan2011, Pei2009}, we will simply use
the one-dimensional search within the interval $[0, T]$  to find the
optimal $\tau$, whose complexity is generally acceptable as known
from \cite{DOA,CFO}.

\subsection{Finding the Solutions}\label{sectionsolution}
The  Lloyd's algorithm is employed here to solve the problem
(\ref{OptP5789}), where local convergence has been proved for some
cases in one-dimensional space. But in general, there is no
guarantee that Lloyd¡¯s algorithm will converge to the global
optimal \cite{Lloyd2008}.
Starting from a feasible solution as the
initial value, e.g., subspaces $\{\Re_{i}\}$ satisfying $p_{i, 0}=\frac{1}{M}$, we repeat the following two steps until the convergence: Step 1) determine the power allocations $\{P_{i}\}$
given  the subspaces $\{\Re_{i}\}$;   Step 2) determine the
subspaces $\{\Re_{i}\}$ given power allocations $\{P_{i}\}$.

 \textbf {Subspaces  Design}: First, we demonstrate
that  the design of the optimal subspace division  $\{\Re_{i}\}$ and
power allocation  $\{P_{i}\}$ is equivalent to
 a modified \emph{distortion measure} design \cite{Lau2008}.
 Incorporating  the power constraints by the Lagrange multipliers $\lambda$ and $\mu$,
we define the following \emph{distortion measure} for optimizing the rate
\begin{align}\label{Prove1}
  R(x, P_{i})=&q_{0}\textup{log}_{2}\left(1+\frac{P_{i}h}{N_{0}}\right)f(x|H_{0})-\mu q_{1}\gamma P_{i}f(x|H_{1})+q_{1}\textup{log}_{2}\left(1+\frac{P_{i}h}{N_{0}+g_{2}P_{p}}\right)f(x|H_{1})\nonumber\\
     &-\lambda P_{i}\left[q_{0}f(x|H_{0})+q_{1}f(x|H_{1})\right].
\end{align}

The optimization problem in (\ref{OptP5789}) is equivalent to selecting
$\{\Re_{i}\}$ and $\{P_{i}\}$ to maximize  the  \emph{average distortion} given by
\begin{equation}\label{Prove2}
    R=\frac{T-\tau}{T}\sum_{i=1}^{M}\int_{x\in \Re_{i}} R(x, P_{i})dx.
\end{equation}

The optimal subspaces $\{\Re_{i}\}$ are then determined
by the \emph{farthest   neighbor rule} as
\begin{equation}\label{Prove432}
\Re_{i}=\{x:~R(x, P_{i})\geq R(x, P_{k}), ~\forall k\neq i\}.
\end{equation}
The following lemma is instrumental to deriving  the optimal
subspaces $\{\Re_{i}\}$.

\emph{Lemma 1}:  For  $x_{1}<x_{2}<x_{3}$, if $x_{1}\in \Re_{i}$,
$x_{2}\in \Re_{k}$ and $i\neq k$, then
$x_{3}\notin\Re_{i}$ must hold.

\begin{proof}  Define the following function
\begin{equation}\label{Prove2}
\begin{aligned}
 &S_{i, k}(x)=R(x, P_{i})-R(x, P_{k})=\frac{x^{\tau f_{s}-1}e^{-\frac{x}{N_{0}}}}{\Gamma(\tau f_{s})}
  \left[\frac{a_{i, k}}{(N_{0}+g_{2}P_{p})^{\tau f_{s}}}e^{\frac{x g_{2}P_{p}}{N_{0}(N_{0}+g_{2}P_{p})}}+\frac{b_{i, k}}{N_{0}^{\tau f_{s}}}\right],\nonumber
\end{aligned}
\end{equation}
where
 $a_{i,
k}=q_{1}\left[\textup{log}_{2}\left(1+\frac{P_{i}h}{N_{0}+g_{2}P_{p}}\right)-\textup{log}_{2}\left(1+\frac{P_{k}h}{N_{0}+g_{2}P_{p}}\right)\right]-\lambda q_{1}( P_{i}-P_{k})-\mu q_{1}\gamma (P_{i}-P_{k}),$
$b_{i,
k}=q_{0}\left[\textup{log}_{2}\left(1+\frac{P_{i}h}{N_{0}+g_{2}P_{p}}\right)-\textup{log}_{2}\left(1+\frac{P_{k}h}{N_{0}+g_{2}P_{p}}\right)\right]
 -\lambda q_{0}( P_{i}-P_{k}).$
From $x_{1}\in \Re_{i}$, $x_{2}\in \Re_{k}$  and (\ref{Prove432}),
we can get that $S_{i, k}(x_{1})>0$ and $S_{i, k}(x_{2})<0$. In
(\ref{Prove432}),  the sign of $S_{i, k}(x)$ is decided by
$\frac{a_{i, k}}{(N_{0}+g_{2}P_{p})^{\tau f_{s}}}e^{\frac{x
g_{2}P_{p}}{N_{0}(N_{0}+g_{2}P_{p})}}+\frac{b_{i, k}}{N_{0}^{\tau
f_{s}}}$ which is a strictly monotonic function. Thus for any
$x_{3}>x_{2}$,  there are  $S_{i,
k}(x_{3})<0$ and $x_{3}\notin \Re_{i}$.
 \end{proof}

Assuming that the range of  $x$ can be divided into  more than $M$  continuous intervals, we immediately know from the \textit{drawer principle} that more than 2 intervals belong to the same subspace which
   is contradicted with  Lemma 1.  Thus  we can conclude that, the range of  $x$ should be divided into  $M$  continuous intervals.  Define $M+1$ thresholds
$\eta_{0},~\eta_{1},...,\eta_{M}$ with $\eta_{0}=0$,
$\eta_{M}=+\infty$. Thus $\Re_{i}$ corresponds to one of
 $[\eta_{j-1}, \eta_{j}),~j=[1,...,M]$. Based on  Lemma 1,
we can calculate $\eta_{j}$ sequentially and assign $\{\Re_{i}\}$ in  Table
\ref{tabPACMAC357}. The answer of $x_{k}$ that satisfies $S_{i, k}(x_{k})=0$ is given by
\begin{equation}\label{Prree2}
\begin{aligned}
 x_{k}=\frac{N_{0}(N_{0}+g_{2}P_{p})}{g_{2}P_{p}}\cdot \textup{In}\left(\frac{-b_{i, k}(N_{0}+g_{2}P_{p})^{\tau f_{s}}}{a_{i, k}N_{0}^{\tau f_{s}}}\right).
\end{aligned}
\end{equation}

\begin{table}[t]
\centering
 \caption{}\label{tabPACMAC357}
 \begin{tabular}{lcl}\midrule
  ~\mbox{\normalsize Subspaces design for $x$ given $\{P_{i}\}$}\\\hline
  ~\mbox{\normalsize $\blacktriangleright$  Initialize the set $\Theta=\{1,...,M\}$;~Set  $i=\textup{arg}\mathop\textup{max}\limits_{j\in \Theta}R(0, P_{j})$, $\Theta \leftarrow \Theta \backslash  i$}\\
 ~\mbox{\normalsize $\blacktriangleright$  For $l=1:M-1$, do}\\
  ~~~~~\mbox{\normalsize 1)}~\mbox{\normalsize  Calculate $x_{k}$ that satisfies $S_{i, k}(x_{k})=0, k\in \Theta$}\\
 ~~~~~\mbox{\normalsize 2)}~\mbox{\normalsize Set $\eta_{l}=\mathop\textup{min}\limits_{k\in \Theta}x_{k}$.  Assign  $\Re_{i}= [\eta_{l-1}, \eta_{l})$}\\
 ~~~~~\mbox{\normalsize 3)}~\mbox{\normalsize Set $i=\textup{arg}\mathop\textup{min}\limits_{k\in \Theta}x_{k}$, $\Theta \leftarrow \Theta \backslash  i$}\\
  ~\mbox{\normalsize $\blacktriangleright$ End for}\\
   ~\mbox{\normalsize $\blacktriangleright$   Set the last element in $\Theta$
   as $i$, $\Re_{i}=[\eta_{M-1}, \eta_{M})$}\\
  \bottomrule
 \end{tabular}
\end{table}\par

 \textbf {Power Allocation}:  After obtaining the threshold $\eta_i$, the probabilities $p_{i, j}$ in (\ref{OptP5789}) can be explicitly expressed as
\begin{equation}\label{eqjoint8}
   p_{i, j}=\int_{\eta_{i-1}}^{\eta_{i}}f(x|H_{j})dx,~i\in[1,...,M],~ j=0,1.
\end{equation}
 First we write the lagrangian $L(P_{i}, \lambda, \mu)$
for problem (\ref{OptP5789}) under  the constraints (\ref{eq1c32})
and (\ref{eq1c34}) as
\begin{align}\label{OptP572189}
L(P_{i}, \lambda, \mu)=&R+\lambda\left(\bar{P}-\frac{T-\tau}{T}\sum_{i=1}^{M}P_{i}\left[q_{0}p_{i, 0}
     +q_{1}p_{i, 1} \right] \right)+\mu \left(\bar{I}-\frac{T-\tau}{T}\sum_{i=1}^{M}q_{1}\gamma P_{i}p_{i, 1}\right),
\end{align}
where $\lambda,~\mu\geq 0$ are dual variables corresponding to (\ref{eq1c32})
and (\ref{eq1c34}).
The lagrange dual optimization can be formulated as
\begin{align}\label{eqOptband}
\mathop\textup{min}\limits_{\lambda\geq 0,~\mu\geq 0}~~g(\lambda, \mu)\triangleq
\sup_{P_{i}\geq 0}L(P_{i}, \lambda, \mu).
\end{align}

In (\ref{OptP5789}),  $
\frac{\partial^{2}R}{\partial^{2} P_{i}}=-\frac{T-\tau}{T}
\left\{\frac{\textup{log}_{2}(e)q_{0}p_{i, 0}}{(P_{i}+N_{0}/h)^{2}}+
\frac{\textup{log}_{2}(e)q_{1}p_{i, 1}}{(P_{0}+(N_{0}+g_{2}P_{p})/h)^{2}}\right\}<0$,
and $\frac{\partial^{2} R}{\partial P_{i}\partial P_{j}}=0, i\neq j$. Since
the constraints are linear functions,  problem (\ref{OptP5789}) is concave
over $P_{i}$.  Thus the optimal value $P_{i}$ of problem
(\ref{eqOptband}) is equal to that of  (\ref{OptP5789}),
and we can solve  (\ref{eqOptband})
instead of (\ref{OptP5789}).
 From (\ref{eqOptband}), we have to obtain
the supremum of $L(P_{i}, \lambda, \mu)$.  Taking the derivative of $L(P_{i}, \lambda, \mu)$ with respect to
$P_{i}$ leads to
\begin{align}\label{Optrew789}
\frac{\partial L(P_{i}, \lambda, \mu)}{\partial P_{i}}=&
\frac{T-\tau}{T}\left\{\frac{\textup{log}_{2}(e)q_{0}p_{i, 0} }{ P_{i}+N_{0}/h}+
\frac{\textup{log}_{2}(e)q_{1}p_{i, 1} }{ P_{i}+(N_{0}+g_{2}P_{p})/h}-\lambda  \left[q_{0}p_{i, 0} +q_{1}p_{i, 1} \right]
-\mu  q_{1}\gamma p_{i, 1} \right\}.
\end{align}

By setting the above equation to 0 and applying the constraint $P_{i}\geq 0$,  the optimal power allocation
$P_{i}$ for given Lagrange multipliers $\lambda$ and $\mu$ is
computed as
\begin{eqnarray}\label{eqOptP0}
  P_{i}=\left[\frac{A_{i}+\sqrt{\triangle_{i}}}{2}\right]^{+},
\end{eqnarray}
where $[x]^{+}$ denotes $\textup{max}(0, x)$, and
\begin{align}\label{Optrew789}
A_{i}=&\frac{\textup{log}_{2}(e) \left[q_{0}p_{i, 0} +q_{1}p_{i, 1}
\right]} {\lambda\left[q_{0}p_{i, 0} +q_{1}p_{i, 1} \right]+\mu
q_{1}\gamma p_{i, 1} }-
\frac{2N_{0}+g_{2}P_{p}}{h},\\
\triangle_{i}=&A_{i}^{2}+\frac{4}{h}\left\{\frac{\textup{log}_{2}(e)
\left[q_{0}p_{i, 0} (N_{0}+g_{2}P_{p})+q_{1}p_{i, 1} N_{0}\right]  }
{ \lambda\left[q_{0}p_{i, 0} +q_{1}p_{i, 1} \right]+\mu q_{1}\gamma
p_{i, 1} }-\frac{N_{0}(N_{0}+g_{2}P_{p})}{h} \right\}.
\end{align}

\emph{Proposition 1}: The power allocation functions $P_{i}$ are non-increasing over $i$.

\begin{proof}
    First, from (\ref{eqjoieqwnt8}), we have
    \begin{equation}\label{eqOp4324P0}
       \frac{f(x|H_{1})}{f(x|H_{0})}=e^{\frac{xg_{1} P_{p}}{N_{0}(N_{0}+g_{1} P_{p})}}
       \left(\frac{N_{0}}{N_{0}+g_{1} P_{p}}\right)^{\tau f_{s}},
    \end{equation}
and obviously it is an increasing function over $x$. Through some simple
manipulations, the monotonicity of $A_{i}$
is equivalent to the monotonicity of the following term
\begin{equation}\label{e1q11tP0}
       C_{i}=\frac{1+\frac{q_{1}}{q_{0}}\frac{p(i, 1)}{p(i, 0)}}
       {1+(1+\mu \gamma/\lambda)\frac{q_{1}}{q_{0}}\frac{p(i, 1)}{p(i, 0)}}.
    \end{equation}
From (\ref{eqOp4324P0}), we can get that
\begin{equation}\label{eqOp4324P121}
   \frac{p(i, 1)}{p(i, 0)}>
   \frac{p(i+1, 1)}{p(i+1, 0)},~\forall i.
\end{equation}

Jointly from (\ref{e1q11tP0}) and (\ref{eqOp4324P121}), we know that
$A_{i}$ is a decreasing function over $i$. Similarly, we can also
prove that $\triangle_{i}$ is a decreasing function over $i$. Thus
from (\ref{eqOptP0}), we can conclude that, $P_{i}$ is a
non-increasing function with respect to $i$. \end{proof}

\emph{Remark}: Proposition 1 shows that, at smaller $x$ the
probability of PU  being busy is smaller, so SU can use higher
transmit power to better exploit the primary band; On the other
hand, at the larger $x$, lower transmit power should be used to
prevent harmful interference to PU. Thus the proposed multiple-level power allocation strategy can also be defined on the the probability of the PU being busy.

Subgradient-based methods are used here to find the optimal Lagrange multipliers $\lambda$ and $\mu$,
e.g., the ellipsoid method and the Newton's method \cite{Boyd2003}. The subgradient of  $g(\lambda, \mu)$ is $[C, D]^{T}$, where
\begin{align}
&C=\bar{P}-\frac{T-\tau}{T}\sum_{i=1}^{M}\bar{P}_{i}\left[q_{0}p_{i, 0}
     +q_{1}p_{i, 1} \right],\nonumber\\
&D=\bar{I}-\frac{T-\tau}{T}\sum_{i=1}^{M}q_{1}\gamma \bar{P}_{i} p_{i, 1},
\end{align}
 while
$\bar{P}_{i}$ is the optimal power allocation for fixed
$\lambda$ and $\mu$ \cite{book2011}. Finally, we summarize the algorithm that computes the  sensing time and multiple-level power allocations in Tab.\ref{tab2}.

\emph{Remark 2}: All computations are performed offline and the resulting power control rule is stored in a look-up table for real-time implementation. Thus the computational complexity
is not significant.
\begin{table}[h]
 \caption{}\label{tab2}
 \centering \begin{tabular}{lcl}\toprule
  \mbox{\normalsize Sensing time and multiple-level power allocations}\\
\toprule
  ~\mbox{\normalsize $\blacktriangleright$~For each $\tau$ in $[0, T]$, do}\\
  ~~~~~\mbox{\normalsize 1)}~\mbox{\normalsize Initialize $\lambda$,
   $\mu$, $\eta_{i}$ satisfying $p_{i, 0}=\frac{1}{M}$}\\
  ~~~~~\mbox{\normalsize 2)}~\mbox{\normalsize  Repeat until $\{\Re_{i}\}$ converge:}\\
    ~~~~~~~~~\mbox{\normalsize - Get $\{P_{i}\}$ using (\ref{eqOptP0});}
\mbox{\normalsize Update $\lambda$ and $\mu$ using the}\\
~~~~~~~~~~~\mbox{\normalsize   subgradient-based method; Until $\lambda$ and $\mu$ converge}\\
    ~~~~~~~~~\mbox{\normalsize -  Update $\{\Re_{i}\}$
 using Table \ref{tabPACMAC357} }\\
  ~\mbox{\normalsize $\blacktriangleright$~ End for}\\
  ~\mbox{\normalsize $\blacktriangleright$~ Optimal parameters: $\tau^{*}=
  \mathop\textup{arg max}\limits_{\tau } R(\tau,
P_{i}, \Re_{i})$,~$\left(P_{i}^{*}, \Re_{i}^{*}\right)=
 \left(P_{i}, \Re_{i}\right)|_{\tau=\tau^{*}}$}\\
  \bottomrule
 \end{tabular}\end{table}

\section{Simulation results}
   In this section, simulations are performed  to evaluate the proposed multiple-level power
allocation strategy in a CR system where the system parameters similar to the references \cite{Fan2011, Kang20092,Pei2009}
are used.
 The frame duration
is taken as $T=100$ ms and the sampling frequency $f_{s}= 1$ MHz. The target detection probability is set to $0.9$ in the opportunistic spectrum access scheme. We
set $g_{1}=N_{0}=0$ dB, $q_{0}=0.7$, $\bar{I}=P_{p}=0.5$, $\bar{P}=10$
dB, $\gamma=h=g_{2}=0$ dB, and unless
otherwise mentioned.

\begin{figure}[h]
\centering
\includegraphics[width=115mm]{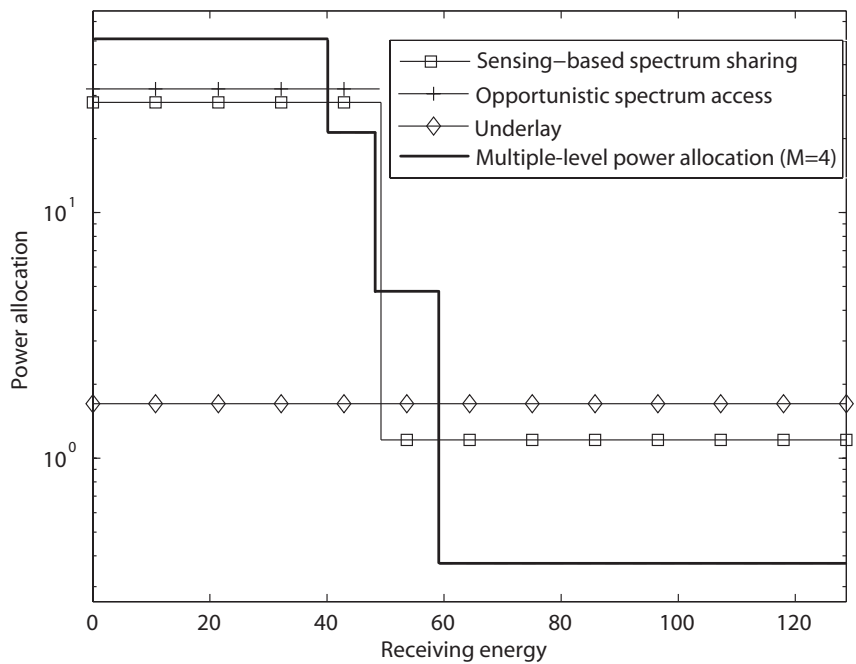}
\caption{Power allocations under the conventional  and  proposed
strategies.} \label{Powerallocation}\vspace{-3mm}
\end{figure}

  Fig. \ref{Powerallocation} compares the power allocations
under the conventional strategies as well as  the proposed one. The figure
shows that  $P_{i}$ for the proposed strategy is a non-increasing
function of the received signal energy.  When $x$ is small, the
proposed strategy allocates more power than the conventional ones,
while when $x$ is large, it allocates less power, thus the average transmit powers are the same for all the strategies.

\begin{figure}[h]
\centering
\includegraphics[width=115mm]{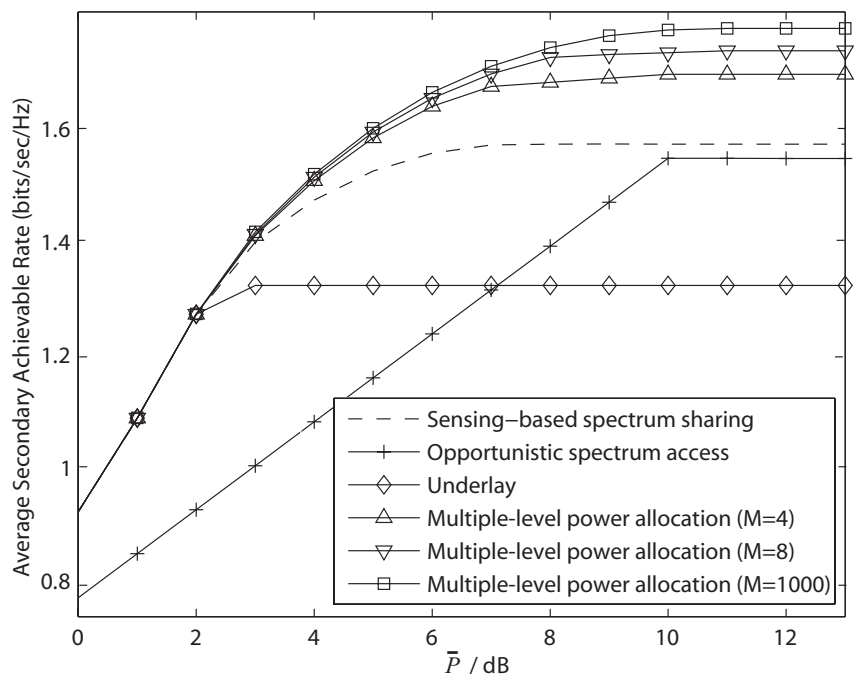}
\caption{Secondary achievable rate vs. $\bar{P}$.}
\label{figaa}\vspace{-3mm}
\end{figure}

Fig. \ref{figaa} shows the average secondary achievable rate.
In the low $\bar{P} $ region,
the proposed strategy and the conventional ones  have the same
rates. However, when $\bar{P} $ is high, the proposed strategy achieves much higher rates. The
rates of all strategies flatten out when $\bar{P} $ is sufficiently
large since the rate is decided by $\bar{I}$ under this condition.
When $M$
increases, the rate of the proposed strategy becomes larger, but the
gain does not improve much when $M$ is large. As $M$ becomes extremely
large, say $M=1000$ in the figure, the rate approaches an upper
limit.  In practice,  we can choose the right $M$ to tradeoff the system complexity and performance,
and in this example $M=4$ serves as a good choice.

\section{Conclusions}
  In this paper, we propose a new multiple-level power allocation strategy for SU
in a CR system.  The
receiving signal energy from PU is divided into different categories
and SU transmits with different power for each category. It is
known that the conventional CR strategies are special cases of the
proposed one. The power levels at SU are obtained by
maximizing the average achievable rate under the
constraints of the average transmit power at SU and the average
interference power at PU. Compared with the conventional power
allocation strategies, the proposed scheme offers significant rate
improvement for SU.

\end{document}